\documentclass[fleqn,10pt]{wlscirep}
\usepackage[utf8]{inputenc}
\usepackage[T1]{fontenc}

\usepackage{color}
\usepackage[normalem]{ulem}

\title{Entropic measure unveils country competitiveness and product specialization in the World trade web}

\author[1]{Gianluca Teza}
\author[2]{Michele Caraglio}
\author[3,4,*]{Attilio L. Stella}
\affil[1]{Department of Physics of Complex Systems, Weizmann Institute of Science, Rehovot 7610001, Israel}
\affil[2]{Institut für Theoretische Physik, Universität Innsbruck, Technikerstraße 21A, A-6020}
\affil[3]{Dipartimento di Fisica e Astronomia  Universit\`a di Padova, Via Marzolo 8, I-35131 Padova, Italy}
\affil[4]{Sezione INFN Universit\`a di Padova Via Marzolo 8, I-35131 Padova, Italy}

\affil[*]{stella@pd.infn.it}


\begin{abstract}
We show how the Shannon entropy function can be used as a basis to set up complexity measures weighting the economic efficiency of countries and the specialization of products beyond bare diversification. 
This entropy function guarantees the existence of a fixed point which is rapidly reached by an iterative scheme converging to our self-consistent measures.
Our approach naturally allows to decompose into inter-sectorial and intra-sectorial contributions the country competitivity measure if products are partitioned into larger categories. 
Besides outlining the technical features and advantages of the method, we describe a wide range of results arising from the analysis of the obtained rankings and we benchmark these observations against those established with other economical parameters.
These comparisons allow to partition countries and products into various main typologies, with well-revealed characterizing features.
Our methods have wide applicability to general problems of ranking in bipartite networks.
\end{abstract}
\begin{document}

\flushbottom
\maketitle
%
%
\thispagestyle{empty}

\section{Introduction}

One of the aims of the economic complexity approach~\cite{Hidalgo2009,Tacchella2012,hausmann2014atlas,Caraglio2016,Teza2018,Teza2018b} is that of establishing consistent rankings of countries in terms of productive efficiency and of products in terms of sophistication~\cite{Helpman1991,Howitt:1998}, with the final goal of estimating both measurable and hidden growth potentials of the economies.
These rankings should result from an analysis of bipartite networks which connect the countries to their products. 
The yearly amounts of the various exports of each country offer the possibility to construct such networks and, on their basis, various algorithms have been proposed so far to obtain measures on which to base the rankings~\cite{Hidalgo2009,hausmann2014atlas,Tacchella2012}. 

While the general ideas inspiring these methods appear quite convincing and promising,
the status of the art regarding their interpretation and implementation is not
completely satisfactory, in spite of the time elapsed since the first proposals.
The mathematical structure of the approach of Ref.~\cite{Hidalgo2009} poses problems of interpretation~\cite{Kemp2014}, while the methods of Refs.~\cite{Tacchella2012} have been shown to suffer poor convergence~\cite{Morrison2017} requiring substantial reformulations~\cite{Servedio2018}.
In addition, most implementations so far make use of binarizations of the bipartite network linking countries to products.
This binarization, although based on RCA~\cite{Balassa1965} criteria, makes only distinction between countries which export and countries which do not export a given product, hence causing a substantial loss of the information provided by the available data concerning exports.
Therefore it would be advisable to construct and use bipartite networks that make full use of the information contained in the original data. 

On a more general level, one should note that, with a single exception~\cite{Mealyeaau2019}, all approaches to economic complexity proposed so far emphasize the importance of diversity as an essential ingredient of the measures~\cite{Hidalgo2009,Tacchella2012,Teza2018}.
In various contexts, ranging from ecology~\cite{Spellerberg2003} to economics itself~\cite{Jacquemin1979,Saviotti2008}, the entropy function first introduced by Boltzmann in statistical physics and later by Shannon in communication theory~\cite{Shannon1948,Weaver1949} is universally used and appreciated as an indicator of diversity~\cite{Jost2006}. 
It is thus somehow surprising that the use of the Shannon entropy function for the construction of economic complexity measures was proposed only in one case so far~\cite{Teza2018}.

In this report, following lines already drawn in Refs.~\cite{Teza2018,Teza2020b}, we show that diversity, as recorded through the Shannon entropy function, can be taken as a basis for establishing meaningful rankings which are fully consistent with the aims of the economic complexity approach, and exempt from difficulties of both interpretation and implementation. 
To initially qualify the bare diversity of export baskets or the ubiquity of productions no distinction is made among products and among producing countries: only monetary amounts are counted, irrespective of the specific export or exporter they refer to.
This corresponds to the standard use of the entropy function in the above mentioned fields of application. 
Distinction among productions and countries is then obtained here by reweighting self-consistently, in terms of the searched measures, the amounts corresponding to products and countries when evaluating the entropy functions.
Such self-consistency can be achieved by running an iterative procedure in which bare entropies enter as starting inputs.
The final measures searched are then obtained as fixed points of the iterative scheme.
While iteration to reach self-consistency is common to all methods proposed so far to obtain complexity measures, in our case its realization within a scheme dictated by the Shannon function guarantees the existence of a fixed point. This feature, and the fact that the iterations reveal the character of contractions rapidly converging to the fixed point, represents an advantage with respect to previous alternative approaches.

The fact that our measure of country competitiveness is expressed as a Shannon entropy function of the weighted percentual shares of the exports in the basket offers also the unique possibility to decompose the same measure into intra- and inter-sectorial contributions, once the shares are summed to represent different sectors of production. 
This is allowed by a key, characterizing property of the function~\cite{Shannon1948} and opens the possibility to revisit in the spirit of the economic complexity approach some methods of analysis already in use in development economics~\cite{Saviotti2008}.

Another virtue of our approach is the wide potential applicability, mainly due to the fact that it is based on the universally used Shannon function.
It can indeed be used in any problem involving (weighted or not weighted) bipartite networks~\cite{Guillaume2004,Newman2010}, when the searched rankings depend in a simple way on the number and strength of the links connecting the two network's layers.
Possible contexts of application range from ecology \cite{Dormann2009}, e.g. the case of plant-pollinator~\cite{Chacoff2012} and predator-pray~\cite{Allesina2008} networks, to collaboration networks~\cite{Ramasco2004} as well as gene sharing networks in microbial communities~\cite{Corel2018}. 
    
This article is organized as follows.
In Section 2 we describe how the data on the exports are used for setting up our bipartite network and provide empirical evidence in favor of the use of the Shannon function.
Section 3 is devoted to the exposition of the iterative algorithm, while Section 4 presents the main results. Section 5 illustrates an example of intra- and inter-sectorial analysis. Conclusions are in Section 6.

\section{Data and bare entropic measures}

\begin{figure}[t]
\centering
\includegraphics[width=\linewidth]{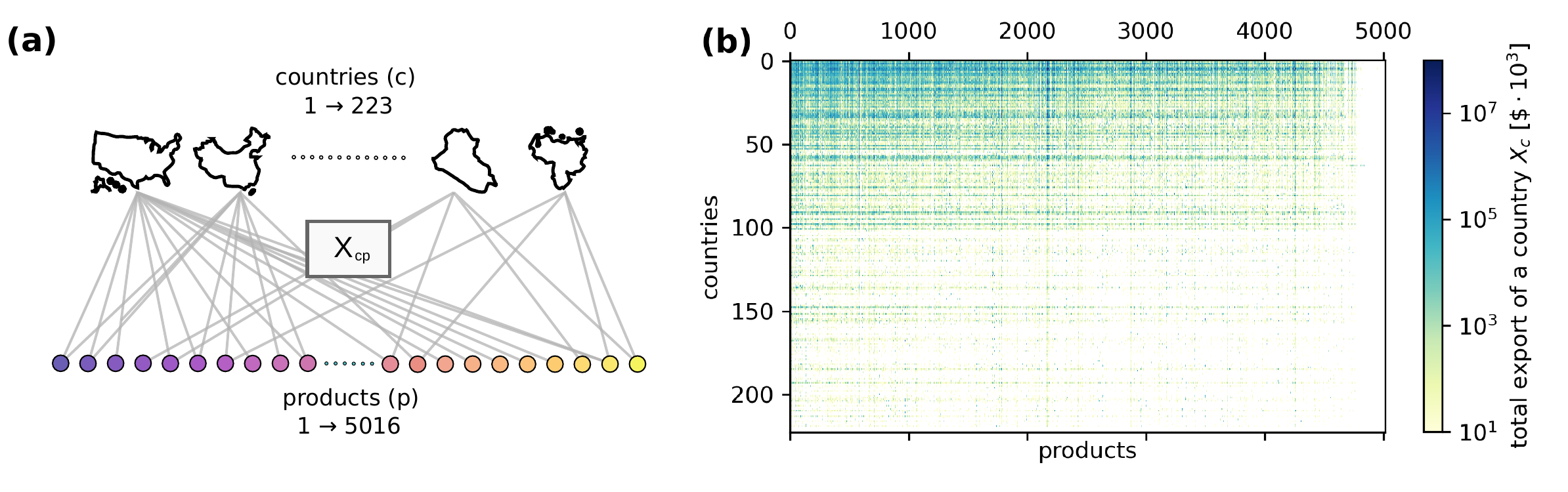}
\caption{\label{fig:bipart} (\textbf{a}) Sketch of the bipartite network associated to the weighted bi-adjacency matrix $X_{cp}$. Every link represents the overall amount (in thousands of current US dollars) of product $p$ exported by a country $c$ towards the rest of the World in a given year. (\textbf{b}) $X_{cp}$ matrix. Countries (rows) are sorted from top to bottom with decreasing diversity $H_c$, while products (columns) are sorted from left to right with decreasing ubiquity $H_p$. Sorting according to these criteria highlights a nested structure of the bipartite network.}
\end{figure}

The data we use for the analysis are extracted from the BACI database~\cite{CEPII:2010-23}, which is a refined version of the freely accessible COMTRADE database~\cite{Comtrade2017}.
This contains export data on a country-to-country level covering 21 years between 1995 and 2015.
The total number of countries contained in the database is $N_c=223$ while products are classified in $5016$ categories according to the Harmonized System 2007 (HS07)~\cite{HS2007}, which consists in a 6 digit code of hierarchical nomenclature.
Every digit represents a category by which the good is classified, and this category becomes more specific as the number of digits increases.
This hierarchical structure consents to naturally coarse-grain the data at different levels, aggregating products sharing the most significant digits.
Being not limited by problems of convergence, our analysis will make use of the full information contained in the original dataset by considering all the 6 digits that characterize the HS07 classification with a total of $N_p=5016$ different product categories.
It is precisely the aggregation represented by the four-digit classification with respect to the six-digit one that we will consider below in order to illustrate the decomposition into inter- and intra-sectorial contributions.  

A preliminary aggregation of the data is carried over the importing countries, in order to obtain the bi-adjacency matrix $X_{cp}(n)$, representing the total amount of product $p$ exported by a country $c$ in a given year $n$ (Fig.~\ref{fig:bipart}). 
Since the analysis does not mix data from different years, from now on we will drop the year index $n$ and specify the year considered in the text.

The starting point for the construction of our measure is a bare Shannon entropy of the nodes of each layer. Given a set of probabilities $\{p_i\}_{i=1\dots N}$, $\sum_i p_i =1$, the Shannon function is defined as
\begin{equation}\label{eq:shann_ent}
H=-\sum_{i=0}^N p_i \log(p_i)
\end{equation}
If we think the $p_i$'s as relative occupations of a collection of $N$ states, Eq.1 expresses the diversity of the corresponding distribution.  Indeed, $H$ increases with both the total number of states or entries, $N$, and with the evenness of the distribution of the $p_i$-s. 
For any $N>1$, $H$ is bounded in the region $[0,\log(N)]$, with the maximum value being reached in the case of exact equipartition $p_i=1/N$ and the minimum when the occupation is concentrated in one single state.

\begin{figure}[t]
\centering
\includegraphics[width=\linewidth]{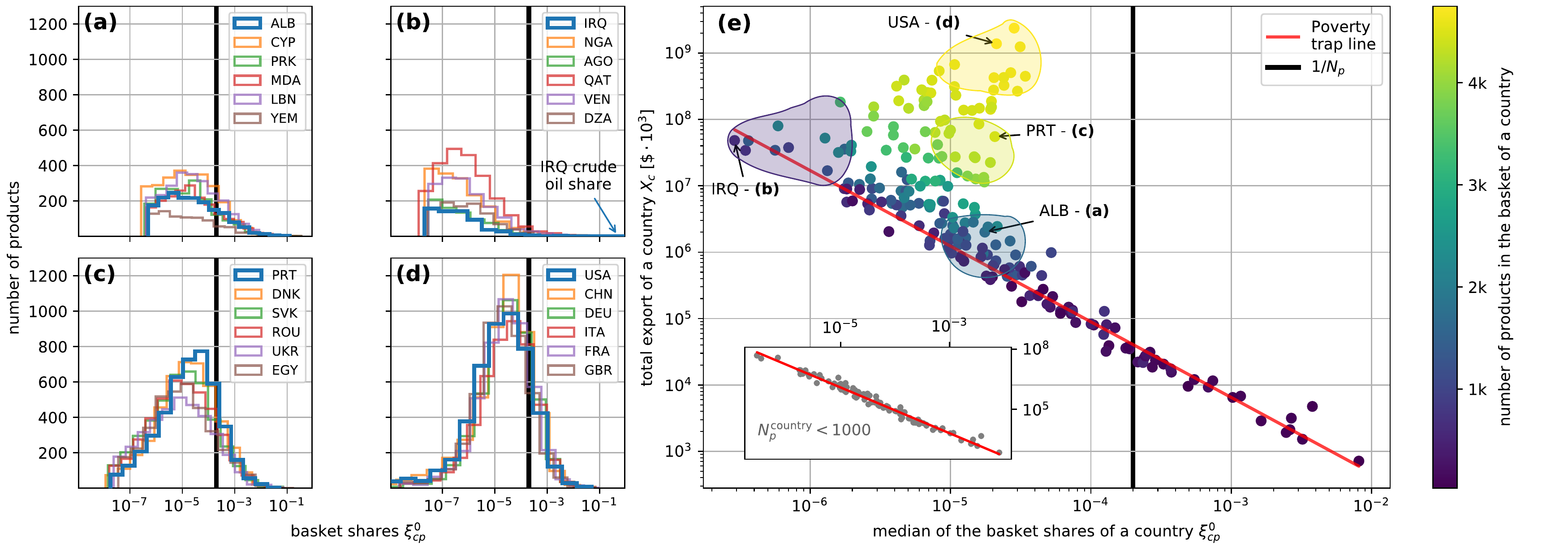}
\caption{\label{fig:bare_ent} (\textbf{a-d}) Examples of histograms of the 2015 basket shares $\xi^0_{cp}$ for different countries grouped among underdeveloped (\textbf{a}), oil dependent (\textbf{b}), semi-developed (\textbf{c}) and fully developed (\textbf{d}).
Albania (ALB), Iraq (IRQ), Portugal (PRT) and the USA will serve as representative examples of these categories throughout the text (country codes follow ISO-3166 standard \cite{ISO3166}). (\textbf{e}) Median of the distribution of the shares vs. total export of a country, colored with respect to the number of exported products. The colored areas show the locations of the four groups of countries whose histograms are reported in Fig. (\textbf{a-d}).
In the inset we highlight the countries exporting less than 1000 products defining a "poverty trap" area. The red line is a linear regression performed over such countries in log-log scale.}
\end{figure}

In the context of the countries-products bipartite network the shares of products in the basket of a given country $c$ can be interpreted as a set of probabilities which altogether provide a measure of the diversification of a country's exports~\cite{Teza2018}. 
Something similar holds also for the products: defining for each country a share of an exported product with respect to the worldwide export of that same product $p$ one can represent somehow the ubiquity of that export~\cite{Teza2018,Teza2020b}. 
Similar uses of the Shannon function to estimate diversification are not new in fields like development economics~\cite{Jacquemin1979,Saviotti2008}. 

So, the shares $\xi^0_{cp} = X_{cp} / \sum_{p'}X_{cp'}$ of the products $p$ in the basket of country $c$ can be plugged in Eq. \ref{eq:shann_ent}, providing us with a "bare" entropic indicator of productive diversification:
\begin{equation}\label{eq:H0c}
H^0_c = -\sum_{p=1}^{N_p} \xi^0_{cp} \log \left( \xi^0_{cp} \right)
\end{equation}
In this sum there are of course terms which are zero corresponding to products which do not appear in the basket of the considered country. 
It is the number of nonzero terms which corresponds to the $N$-dependence we were discussing in connection with Eq.~\ref{eq:shann_ent}.
In Fig.~\ref{fig:bare_ent}a-d we present histograms of the basket shares $\xi_{cp}^{0}$ of 2015 in the case of countries included in the four different regions
highlighted in Fig.~\ref{fig:bare_ent}e. Countries belonging to the same area exhibit qualitatively similar distributions, regardless of the different number of
products exported (histograms are not normalized). Remarkably, we find that such a qualitative classification through the basket shares
distributions allows to characterize countries in categories in line with the economic narrative: Fig.~\ref{fig:bare_ent}a reports histograms of underdeveloped
countries, including Albania (ALB); Fig.~\ref{fig:bare_ent}b of oil dependent economies, including Iraq (IRQ); Fig.~\ref{fig:bare_ent}c of moderately developed economies, including
Portugal (PRT), and Fig.~\ref{fig:bare_ent}d of fully developed nations, including USA. The different total numbers of products exported by the various countries
are given by the areas under the corresponding histograms. Up to differences in these numbers the shapes of the histograms are rather similar
within each group. Therefore, throughout the rest of the paper, USA, IRQ, PRT and ALB will serve as representative examples for countries
characterized by these differently structured economies.
We also notice that fully and moderately developed countries like USA and PRT present a relatively narrow peak rather close to the equipartition value of the full range of products $1/N_p$, while both IRQ and ALB show a broader peak, with IRQ's peak being much further away from this equipartition value.
Moreover, while for example PRT and IRQ share a very similar total export $X_c= \sum_p X_{cp} \sim 5 \cdot 10^{10} \$ $, we can clearly see how the distribution are very different: not only the number of exported products (area of the histogram) is much larger for PRT, but also the distribution of basket shares in the case of IRQ is extremely uneven. 
This is due to the overwhelming dominance of the few oil related exports (highlighted with a blue arrow in Fig.~\ref{fig:bare_ent}b), with total shares holding 98\%. 

One parameter that also captures and partly quantifies the above considerations is the median of the distributions of the basket shares.
With the due precautions connected to the fact that the median of the distribution is also depending on the number of products exported by the country and thus cannot be naively compared to the world equipartition value $1/N_p$,
this parameter provides an insight complementary to that of the total export.
In Fig.~\ref{fig:bare_ent}e we show how such quantity relates with the total export $X_c$ for each one of the 223 countries of the data-set, always in the year 2015.
The color scheme used in Fig.~\ref{fig:bare_ent}e reflects the number of products exported by each country.
It is remarkable how all the countries exporting less than 1000 products are found to lie on a line, exhibiting a very high correlation between the overall export of a country and the median of its basket shares.
Such behavior defines a sort of "poverty trap" on which are lying the countries whose export depends only on a blind exploitation of the natural resources at disposal. 
Note that, in spite of its relatively high total export, also Iraq is lying on this line because of its complete dependence on oil exports.
Countries that managed to improve their overall economic wealth had to increase both the number of products and the diversification of the total production. 
This strategy allowed them to detach from the "poverty trap" line in a direction which brought them closer to the ideal equipartition value $1/N_p$.

The above considerations point to basket shares distributions as a key ingredient for the construction of a measure for the productive efficiency of the countries. At the same time,
the Shannon entropy function introduced in Eq.~\ref{eq:H0c} qualifies as a natural candidate to represent a single value quantifier of all the "information" contained in such distributions.

Similarly, one can think of an analogous argument also for the layer of the products~\cite{Teza2018}. 
For any given product $p$ we can define an ubiquity measure that takes into account both the number of countries that are able to export such product and how evenly its offer on the global market is distributed. 
To do so, we define an \emph{export} share $\zeta^0_{cp} = X_{cp} / \sum_{c'}X_{c'p}$ which is normalized with respect to the overall export of that same product at a global level. 
The bare entropic indicator for the ubiquity of products will therefore be:
\begin{equation}\label{eq:H0p}
H^0_p = -\sum_{c=1}^{N_c} \zeta^0_{cp} \log \left( \zeta^0_{cp} \right)
\end{equation}
These quantifiers will be at the basis of our measure constructions. 

\section{Iterative scheme}

The entropic bare measures of Eqs.~\ref{eq:H0c} and~\ref{eq:H0p}, although already rather sound, are not making full use of the information at our disposal.
For instance, the formulation of $H^0_c$ is indifferent to interchanges among products, meaning that swapping the exports of two products would leave the entropy unaltered.
Analogously, swapping the exports of two countries for the same product $p$ would preserve the associated ubiquity $H^0_p$.
Nevertheless, each product (and country) should enter differently in the evaluation of the measures. 
This is also required by the basic assumptions of the economic complexity approach, which through a comparative analysis aims at distinguishing among products and among producing countries. 
In general, one would intuitively expect "important" products to weight more in a country's wealth indicator. 
Such importance should be determined by how many developed countries are exporting these products.
We can introduce such dependencies by reweighing the shares entering Eqs.~\ref{eq:H0c} and~\ref{eq:H0p} with weights self-consistently related to diversities and ubiquities.
The imposition of such self-consistency can be achieved by the construction of an iterative algorithm in which at each step we evaluate finer measures of diversity and ubiquity by using weights determined by the same quantities evaluated at the previous step.
A general formulation of such algorithm reads:
\begin{equation}\label{eq:iter_scheme}
\begin{cases}
H_c^{(k+1)} &= -\sum_{p} \xi^{(k)}_{cp}
\log \left( \xi^{(k)}_{cp} \right) \\
H_p^{(k+1)} &= -\sum _{c} \zeta^{(k)}_{cp}
\log \left( \zeta^{(k)}_{cp} \right)
\end{cases}
\end{equation}
with shares at the $k$-th step defined as
\begin{equation}\label{eq:iter_scheme2}
\xi^{(k)}_{cp}=\frac{X_{cp}f\left(H_p^{(k)}\right)}{\sum_{p'}X_{cp'}f\left(H_{p'}^{(k)}\right)} \ ;\ 
\zeta^{(k)}_{cp}=\frac{X_{cp}g\left(H_c^{(k)}\right)}{\sum_{c'}X_{c'p}g\left(H_{c'}^{(k)}\right)}
\end{equation}
Here $f$ and $g$ are two functions that, respectively, take in as argument $H_p$ and $H_c$.
Inspecting Eq.s~\ref{eq:iter_scheme} and~\ref{eq:iter_scheme2} one can easily appreciate the role played by the weights introduced to modify the bare shares in the Shannon function.
For the sake of argument and to simplify notation, let us look at the $f$'s and the $g$'s in Eq.~\ref{eq:iter_scheme2} as to independent variables with labels referring to the corresponding product $p$ ($f(H^{(k)}_p) \to f_p$) or country $c$ ($g(H^{(k)}_c) \to g_c$).
Then, let us consider for example a country with a very special basket having one single dominant product $p$ ($X_{cp} \simeq 1$) and all the other products $p' \neq p$ with $X_{cp'} \simeq 0$.
Due to the shape of the function $-x \log(x)$ in the interval $[0,1]$, an increase of $f_p$ would lead to a decrease of $H^{(k+1)}_c$ both because the weighted importance of product $p$ increases (increase of $\xi^{(k)}_{cp}$ towards $1$) and because the weighted importance of the other products $p'$  decreases ( $\xi^{(k)}_{cp'}$ become closer to $0$).
On the contrary, a decrease of $f_p$ would lead to an increase of $H^{(k+1)}_c$.
So, an unbalanced basket dominated by a single export becomes even less efficient if that export is further valued. 
Thus, the logic at the basis of using weighted shares is such to allow distinction among products and to favor an optimal balance among all productions at the same time.

The choice of the functions $f$ and $g$  in Eq.~\ref{eq:iter_scheme} is clearly not unique, and in Ref.~\cite{Teza2018} a particular form of self-consistency was preliminary proposed. 
In any case the choice is dictated by the aims of the analysis. 
Thus, it needs to reflect the fact that the more a product is ubiquitous the less it should contribute to the diversity of a basket of a country.
Also, the more a country is wealthy and developed, the less it should count in determining the ubiquity of a product.
Therefore we need to define two functions $f$ and $g$ that invert the concept of diversity and ubiquity, respectively.
Thanks to the boundedness of the Shannon entropy function, a simple and convenient way to achieve such inversion is through the following simple linear relations:
\begin{eqnarray}\nonumber
f(H_p) &= \log(N_c) - H_p \\
g(H_c) &= \log(N_p) - H_c
\end{eqnarray}
These positive weights, with their continuity, grant existence of a fixed point~\cite{Kellogg1976} and in fact stability of the iterative algorithm.

\begin{figure}[t]
\centering
\includegraphics[width=\linewidth]{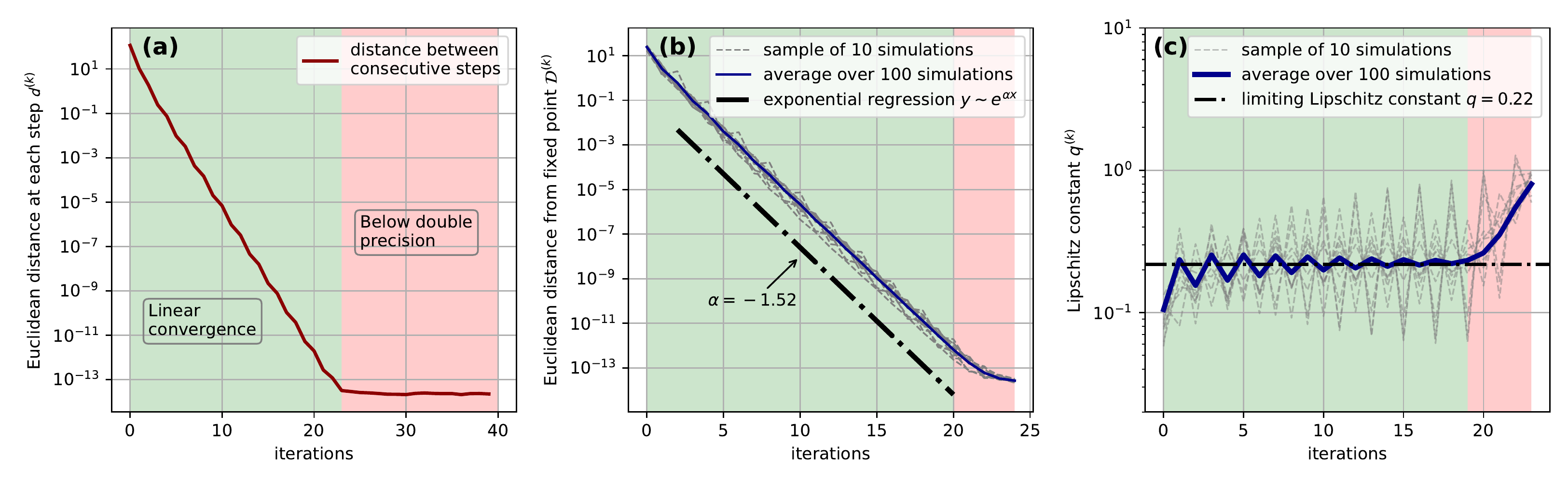}
\caption{\label{fig:convergence} Convergence of the iterative scheme (\textbf{a}) Euclidean distance $d^{(k)}$ evaluated between consecutive steps of the iterative algorithm of Eq. \ref{eq:iter_scheme}. 
It clearly exhibits linearly fast convergence (green area of the plot), while after 23 iterations double precision limit is reached (red zone in the plot). (\textbf{b}) Euclidean distance $\mathcal{D}^{(k)}$ from the fixed point of 100 iterations initiated with random initial conditions. 
Linear convergence to the same fixed point is observed, proving its uniqueness and global convergence of the algorithm. 
(\textbf{c}) Evaluation of the Lipschitz constant $q$ as the ratios of distances from the fixed point of consecutive iterations. A sample of the 100 iterations starting from random initial conditions are colored in grey, whereas in blue we see the average performed over the whole data sample. The dashed black line represents the estimation of the Lipschitz constant obtained by the regression illustrated in the previous panel.}
\end{figure}

Indeed, the scheme of Eq. \ref{eq:iter_scheme} can be seen as a map $\varphi$ of a closed, compact set in itself
\begin{equation}
\varphi: [0,\log N_p]^{N_c} \otimes [0,\log N_c]^{N_p} \rightarrow [0,\log N_p]^{N_c} \otimes [0,\log N_c]^{N_p}
\end{equation}
Such a peculiar feature is guaranteed by the boundedness of the Shannon's entropy function, which (together with the continuity property of $\varphi$ allows us to apply Brower's fixed point theorem \cite{Kellogg1976} to ultimately prove the existence of a fixed point $\{H_c,H_p\}^{c=1\dots N_c}_{p=1\dots N_p}$ for the map $\varphi$.
In Fig. \ref{fig:convergence}a we show indeed how such fixed point is reached exponentially fast by iterating numerically the scheme of Eq. \ref{eq:iter_scheme}. Evaluating the Euclidean distance between two consecutive steps, defined for every $k\in \mathbb{N}$ as:
\begin{equation}\label{eq:euc_dist_step}
d^{(k)}=\left( \sum_c \left( H_c^{(k+1)}-H_c^{(k)} \right)^{2} +
\sum_p \left( H_p^{(k+1)}-H_p^{(k)} \right)^{2} \right)^{1/2}
\end{equation}
Numerically we also tested the uniqueness of such fixed point (to check if the algorithm is globally convergent) by iterating the scheme starting from different randomly chosen initial conditions. We evaluated at each step the distance with respect to the previously obtained fixed point, defined as
\begin{equation}\label{eq:euc_dist_fix}
\mathcal{D}^{(k)}=\sum_c \left( \left(H_c^{(k)}-H_c \right)^{2} +
\sum_p \left( H_p^{(k)}-H_p\right)^{2} \right)^{1/2}
\end{equation}
and found for it an exponential decay with $k\to\infty$: $\mathcal{D}^{(k)}\propto e^{\alpha k}$ with $\alpha=-1.52$ (Fig. \ref{fig:convergence}b). With such parameter we were therefore able to estimate the Lipschitz constant \cite{Evans2015}, defined as $q=\lim_{k \to \infty} \mathcal{D}^{(k+1)}/\mathcal{D}^{(k)}=0.22$ (see Fig. \ref{fig:convergence}c).
These estimates allow to classify the algorithm as globally convergent with a linear rate of convergence ($q < 1$ means that the map associated with the algorithm is a contraction).

The scheme therefore always ultimately converges to the same fixed point for the entropies of countries and ubiquities of products, for which the following consistency relations hold
\begin{equation}\label{eq:H_fix_point}
\begin{cases}
H_c &= -\sum_{p} \xi_{cp}
\log \left( \xi_{cp}  \right) \\
H_p &= -\sum _{c} \zeta_{cp} 
\log \left( \zeta_{cp} \right)
\end{cases}
\end{equation}
where we introduced the weighted shares normalized with respect to a country's export
\begin{equation}\label{eq:H_fix_point_weighted_shared}
\xi_{cp}=\frac{X_{cp}f\left(H_p\right)}{\sum_{p'}X_{cp'}f\left(H_{p'}\right)} \ ;\ 
\zeta_{cp}=\frac{X_{cp}g\left(H_c\right)}{\sum_{c'}X_{c'p}g\left(H_{c'}\right)}
\end{equation}
The iteration procedure, through its fixed point, solves the mathematical problem of providing the solution of the self-consistency conditions in Eqs. \ref{eq:H_fix_point},\ref{eq:H_fix_point_weighted_shared}.

From an economic point of view, the iteration scheme can be interpreted as an attempt to progressively establish the weights of the factors playing a role
in the production network of a good. Trying to establish the relevance of all the intermediate steps that lead to the realization of a product reveals
to be an extremely challenging task. Production chains are indeed the result of multiple complex interactions, often involving products realized
across different countries at a global scale. Our approach naturally overcomes the limitations induced by analyzing a single national value chain:
through exports, all the national productivity chains are effectively interconnected with one another, merging into a global value chain. One
therefore needs to take into account such connections, keeping in mind that every country is nevertheless required to offer a highly diversified
basket of products in order to retain an adequate self-sustainability as well as the ability to compete as a driving force in the global economy. The iterative scheme endogenously captures the intrinsic relevance of both countries and products, establishing weights for these
structural ingredients of the global value chain.

\section{Results}

\begin{figure}[t]
\centering
\includegraphics[width=0.95\linewidth]{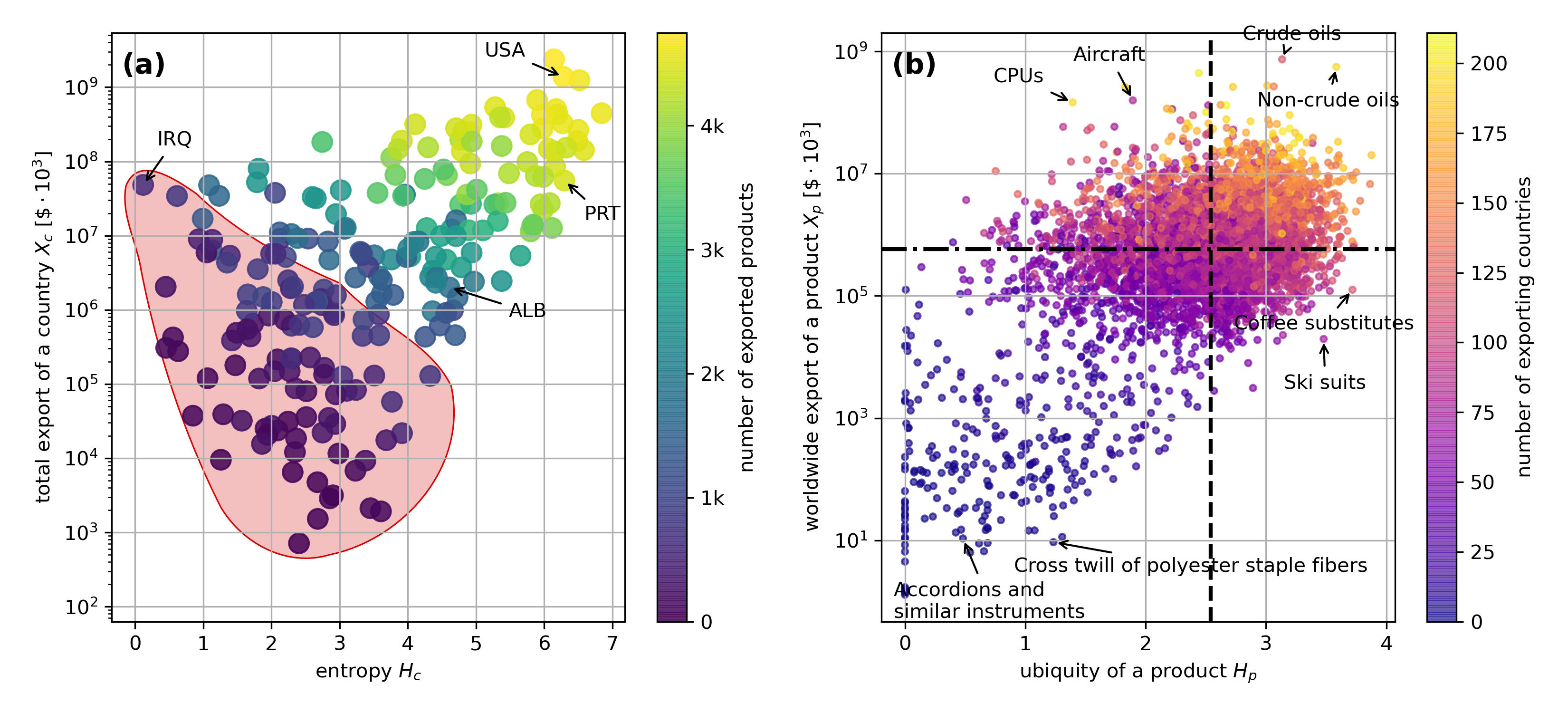}
\caption{\label{fig:resultsH} (\textbf{a}) Fixed point of the entropic measure $H_c$ for the countries vs. the overall export $X_c$. 
The colors used are related to the number of exported products. 
We can see how the countries that belong to the "poverty trap" line ($N<1000$) introduced in Fig.~\ref{fig:bare_ent}b are here scattered over a wide area (highlighted in red).
(\textbf{b}) Ubiquity of the products $H_p$ vs. World aggregated export of that same product. The color scale reflects the number of countries exporting such product. The black lines mark the medians of the ubiquity and of the amount of worldwide export. They allow us to classify products in 4 different main categories.}
\end{figure}

\begin{figure}[t]
\centering
\includegraphics[width=0.95\linewidth]{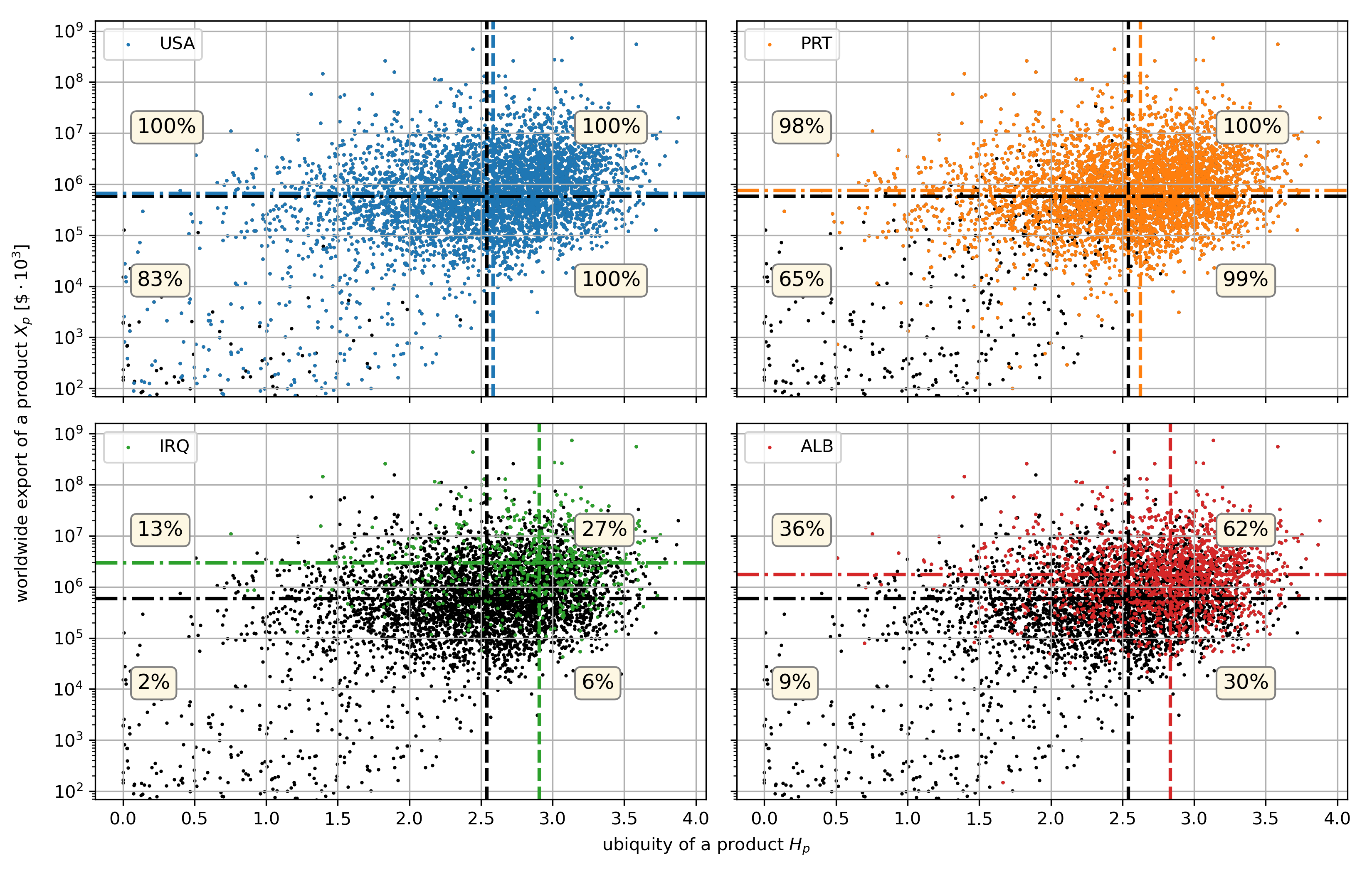}
\caption{\label{fig:results_examples}
    Prouducts exported by USA (top-left), Portugal (top-right), Iraq (bottom-left) and Albania
    (bottom-right) in the worldwide export vs. ubiquity plane of Fig. \ref{fig:resultsH}b.
    Black dots represent all the existing products, while the colored ones are only those actually
    exported by each country. It is clear how least developed countries tend to have medians
    (colored lines) leaning towards more ubiquitous and highly exported products (the black lines
    are the medians for all the existing products).}
\end{figure}

In this section, we present some of the main results emerging from the analysis performed with our algorithm. 

In Fig.~\ref{fig:resultsH}a we report on the horizontal axis our fixed point entropic measures $H_c $ of the 223 countries for the year 2015. 
On the vertical axis is the total export of each country in the same year, and the colors vary according to total number of exports.
Countries with maximal entropic efficiency are in fact those with the highest total export value on the top right corner. 
Not surprisingly, also in view of what we anticipated in the second section (Fig.~\ref{fig:bare_ent}b), the countries in this corner are also those with the largest numbers of productions. 
Here we can identify in particular the four countries pointed out in Fig.~\ref{fig:bare_ent}a-d as representative of four different types of economy. 
While USA and PRT are in the top right corner, although with different levels of total export and number of products, IRQ is in the top left corner, of countries which are based almost exclusively on export of natural resources. 
ALB clearly differentiates from IRQ with its higher entropic measure, largely due to the relatively broader distribution of shares in the basket.
Notably, this entropic measure successfully captures the distance from the "poverty trap" highlighted in Fig.~\ref{fig:bare_ent}b and here (Fig. \ref{fig:resultsH}a) represented by the region in which the countries with less then $1000$ exported products lay. 
The countries in this trap are now spread on a rather large region colored in red.
The entropic measure also provides a better estimator of country efficiency than the median of the bare basket shares distribution, according to which ALB, PRT and USA appear to be equidistant with respect to the equipartition line.

The entropic ubiquities of products are shown in Fig.~\ref{fig:resultsH}b, where they are reported on the horizontal axis. 
The year of reference is again 2015. 
On the vertical axis is reported the worldwide amount of export of each product and the scale of colors is in accordance with the number of exporting countries. 
The horizontal and vertical lines simply represent median values of the corresponding quantities. 
In this way one partitions the products in four categories: in the top right corner fall the products with large ubiquity and with large volume of global export. 
These products are generally exported by a large number of countries, which is often also responsible for the large value of the global exports. 
Products like crude and non-crude oils are also falling in this category.
The products in the top left corner are those with low ubiquity and include computer processor units (CPUs) and aircrafts. 
In this case, while the amount of global export is often very large, the number of exporting countries is small, as appropriate for very specialized productions. 
In the right bottom corner the are exports with high ubiquity exported in variable but moderate total amounts by a number of countries definitely lower than that of nations contributing  to the upper right corner.
Finally, in the lower left corner there are products which are very marginal in the global economy. 
These products have low ubiquity, small total amount of exports, and few exporting countries.

It is of particular interest to find out how specific countries are positioned with their exports compared to the global situation represented in Fig.~\ref{fig:resultsH}b. 
In Figs.~\ref{fig:results_examples}a-d we do this comparison for USA, PRT, IRQ and ALB. 
For instance, in the case of USA we reproduce Fig.~\ref{fig:resultsH}b with all products in black, coloring in blue only those that are actually exported by the USA. 
Remarkably, we see that almost all points are blue, meaning that the USA are exporting almost all products present in the global basket. 
To quantitatively appreciate how the exports of the USA are divided in the four sectors identified in Fig. \ref{fig:resultsH}b, we report (in blue) the medians pertaining to only the blue dots.
The percentages reported in each sector give a measure of the overlap between USA export basket and global basket in that sector. 
For the USA the only percentage lower than $100\%$ is in the lower left corner. 

Also for PRT all the sectors are rather well occupied, but one can already appreciate a slight shift to the right of the horizontal median, indicating an average higher ubiquity of exports with respect to the USA.
In the case of IRQ and ALB we see instead a very sensible shift of the medians, and, especially in the case of IRQ considerably low percentages, indicating that these economies are definitely far from reproducing the global export basket.
The shifts indicate that such countries are forced to export products that are not only on average more ubiquitous, but also more dominant in the world aggregated basket of exports.
This is not surprising, since the  export aggregated at a global level exactly coincides with the worldwide import. This means that poorly developed countries, lacking the possibility to export elaborated products concentrate their efforts in goods that are relatively easy to produce and highly demanded in the global panorama.

A peculiarity that emerges from this analysis is how developed countries seem to export all kinds of products, including those with extremely high ubiquity. Least developed countries, on the other hand, are lacking an enormous percentage of such products. 
This suggests that a country, in order to increase its development, cannot aim to just produce the most complex and/or heavily exported products on the market. 
There are indeed articulated constraints 
linked to the possibility of producing such goods. Such constraints in many cases remain even hidden (related with the so called "intangibles").
This analysis sheds light on how the lack of apparently irrelevant products in the basket might heavily impact the overall development itself of a country down the line.
In other words, we understood how a developed country can hardly refrain from continuing to keep its exports as diverse as possible, including ubiquitous and poorly demanded products.

It is interesting to compare our rankings with those one obtains with the approach of Ref.~\cite{Tacchella2012}. Here the rankings of that approach were obtained by use of the full 6-digit code of the HS7 product classification
For sure the fitness method, at this level of resolution can be anticipated to suffer of convergence problems~\cite{Servedio2018}, which in fact were already encountered at the four digit level~\cite{Morrison2017}.
Instability issues of the algorithm impose to cut somehow arbitrarily the number of iterations in this case.
The problem arises from the implementation of non linearity in the algorithm~\cite{Morrison2017}, which leads some products' measures (therein called \emph{complexity}) to progressively approach the limiting value 0.
Such behavior denotes an instability of the scheme, extremely marked for the complexities but less visible for the fitness measure of countries.
Indeed, evaluating the Spearman correlator~\cite{Hogg:1995} between the countries' fitness $F_c$ resulting after hundred iterations and our diversity $H_c$ of Eq.~\ref{eq:H_fix_point} yields the rather high value $\rho_S\left\{H_c,F_c\right\}=0.94$ (6 digits code of the HS07 classification, year 2015).
The Spearman correlator between products' complexity $Q_p$ and our ubiquity $H_p$ of Eq.~\ref{eq:H_fix_point} yields $\rho_S\left\{H_p,Q_p\right\}=-0.32$.
The minus sign comes from the fact that ubiquity is conceptually an inverse of complexity. 
The remarkable distance of the obtained correlation from the value $-1$ indicates that the two approaches give rather different rankings for the products.

\section{Coarse-grained analysis, and intra- vs inter-sectorial contributions}

\begin{figure}[t]
\centering
\includegraphics[width=0.95\linewidth]{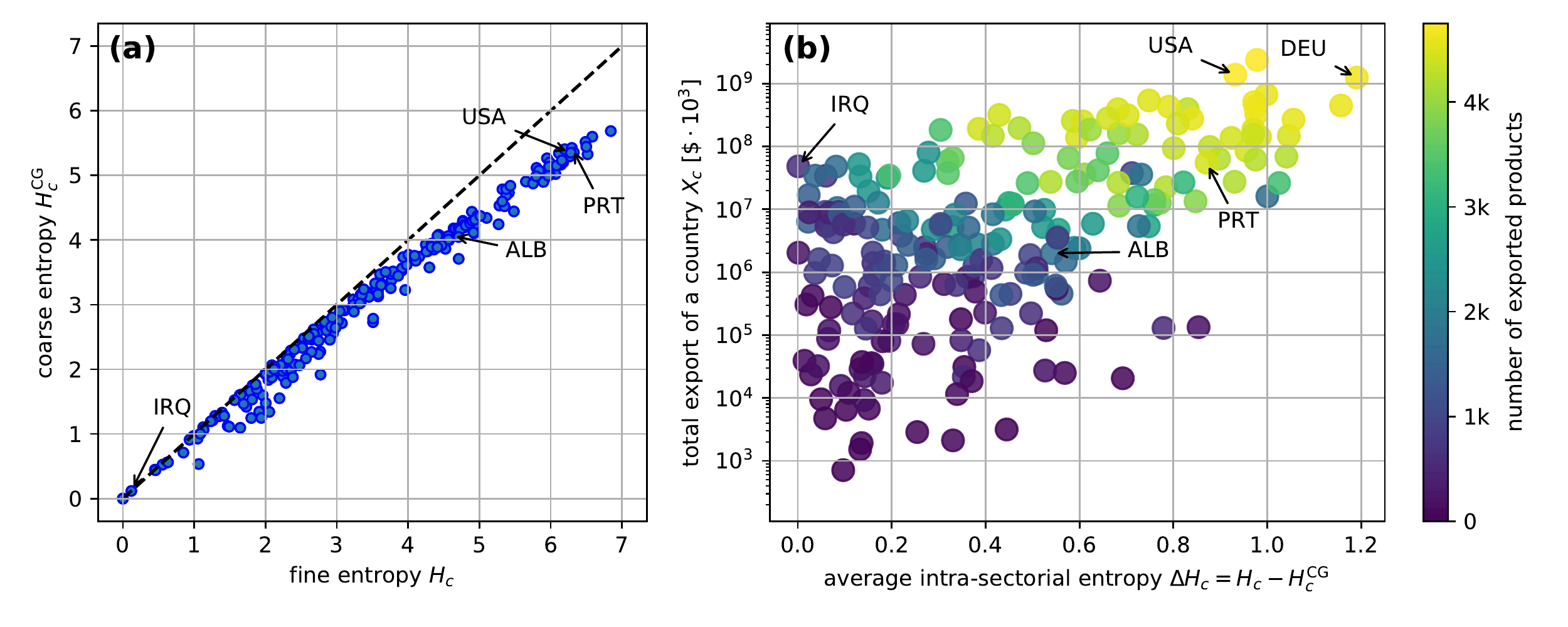}
\caption{\label{fig:CG} (\textbf{a}) Comparison between the fine-grained (6 digit HS07 classification) entropies $H_c$ of the countries for Eq. \ref{eq:H_fix_point} and the coarse-grained (4 digits HS07 classification) entropies $H^{\mathrm{CG}}_c$ of Eq. \ref{eq:HCG_c}. The quadrant bisector marks the inequality $H_c>H^{\mathrm{CG}}_c$ relating the two.
(\textbf{b}) Average intra-sectorial entropy $\Delta H_c$ of Eq. \ref{eq:DeltaH_c} plotted against the overall export of a country $X_c$. Germany (DEU) emerges as the country with the highest $\Delta H_c$, implying an extremely articulated structure of the export shares.
The gradient of the color of every point is related to the overall number of exported products by each country in the 6 digit HS07 classification.}
\end{figure}

The entropy function satisfies a special summation rule when clustering states~\cite{Shannon1948,Jacquemin1979}. 
The HS07 classification of products~\cite{HS2007} is intrinsically structured as a multilayered nomenclature, where the leftmost digits are shared by products belonging to a more general macro-category.
So far our analysis was carried out at the finest possible level of details offered by the dataset, consisting in 6 numerical digits identifying a total of 5016 product categories.
According to the convention of the HS07 classification, one can group together products sharing the same 4 leftmost digits resulting in a total of 1241 macro-categories $P$.
Referring to the fixed point shares $\xi_{cp}$ of Eq.~\ref{eq:H_fix_point} we introduce the coarse-grained shares $\xi_{cP}=\sum_{p\in P}\xi_{cp}$, for which one can evaluate a \emph{coarse-grained} Shannon entropy:
\begin{equation}\label{eq:HCG_c}
H^{\mathrm{CG}}_c = -\sum_P \xi_{cP} \log \left( \xi_{cP} \right)
\end{equation}
One can easily see that the inequality $H_c > H^{\mathrm{CG}}_c $ holds, with the equality holding only in the case in which for every coarse-grained category $P$ there is one and only one fine-grained category $p$.
The difference between the two entropies is indeed related to the average \emph{intra-sectorial} entropy of a macro-category $P$ defined as
\begin{equation}
H_{cP}=-\sum_{p \in P} \frac{\xi_{cp} }{\xi_{cP}} \log \left( \frac{\xi_{cp} }{\xi_{cP}} \right).
\end{equation}
Straightforward calculations show that the difference between the two level entropies is the following weighted average:
\begin{equation}\label{eq:DeltaH_c}
\Delta H_c = H_c-H^{\mathrm{CG}}_c= \sum_P \xi_{cP} H_{cP}.
\end{equation}
This summation rule
allows us to regard the coarse-grained entropy of Eq. \ref{eq:HCG_c} as \emph{inter-sectorial} and $\Delta H_c$ in Eq. \ref{eq:DeltaH_c} as \emph{intra-sectorial} contributions to the total entropic measure of country $c$. 
As we see in Fig.~\ref{fig:CG}a, the intra-sectorial contribution is more substantial for countries with high entropic measure, indicating the importance of organization within sectors.
In Fig.~\ref{fig:CG}b we report also the total export of each country as a function of $\Delta H_c$.
One sees that the countries with the largest export, both in terms of total amount, and in terms of number of products, are those with highest intra-sectorial contribution. 
Germany (DEU) reaches the top of the list in this case.
At the same time, most of the countries belonging to the "poverty trap" show a low $\Delta H_c$, which is another indication of the scarce inner organization of their economic structure.

One can of course perform intra- and inter-secorial decompositions also with reference to larger sectors of the economies. Such decompositions were already considered in the economic literature with reference to bare entropic indicators~\cite{Saviotti2008}. This opens novel perspectives of analysis within the complexity approach.

\section{Conclusions}

In this report we showed how the Shannon entropy function can be used to develop a consistent and rapidly-convergent method for the evaluation of economic complexity measures.
In view of the universally accepted meaning of the Shannon function, our construction gives concrete support to the expectation that diversity can be assumed as a basic ingredient of such measures.

From a mathematical and numerical point of view, the success of our approach relies on the continuity and boundedness of this function, which guarantees both the existence of and the convergence to a fixed point in our iterative scheme.
Here our method was proven to be effective on a weighted bipartite network between two layers composed of 5016 (products) and 223 (countries) nodes, and with the links' weight spanning values across more than 9 decades.
The proven stability of our algorithm therefore acquires additional significance in the perspective of more general applications, giving confidence to obtain consistent and meaningful results also outside the particular context and database considered here.

Results show that, using our refined entropic measure, it is possible to establish a meaningful and unambiguous ranking of countries, with the most developed ones also scoring the highest entropic measures.
On the other hand, countries that have a poorly developed economy or simply rely on the exploitation of their raw materials are characterized by low values of our entropic index.
In a similar fashion, products can be ranked according to their ubiquity. The more ubiquitous is a product, the more it is exported (or imported) in the global market.
Developed countries characterize themselves as exporter of products in the whole range of ubiquities.

Another key mathematical property is the possibility to decompose the function into parts which can be ascribed to intra- and inter-sectorial contributions \cite{Shannon1948,Weaver1949,Teza2020b}.
This gives the possibility to quantitatively analyze the complexity measure of countries in terms of the interplay among different categories in which one may wish to partition the production, and in terms of the the importance of each individual category.
It emerges that developed countries are the ones displaying the largest intra-sectorial entropies.

While the analysis presented here referred only to a particular year of the dataset, a comparison of the results for different years opens to an analysis of the evolution dynamics of the complexity measures.
This would allow to argue further information both on the different growth potential of each country and on the evolution of the hierarchy of importance of various products in the global economy.
A preliminar analysis along these lines~\cite{Teza2018,Teza2020b} has already highlighted a strong bond connecting the entropic diversity of a country and the hierarchy of products to a dynamical model regulating the yearly evolution of single exports for every country.
For the time evolution of the basket compositions, the present authors have recently developed a statistical mechanics model~\cite{Caraglio2016,Teza2018,Teza2018b}, calibrated on the same database used here, which clarifies the non-equilibrium character of its dynamics.
The combined use of this model with the entropic complexity analysis offers in perspective possibilities of predictions going beyond standard regression analysis.
As we showed in Section 5, the entropic measures allow to keep track of the whole information of the most detailed data, without dispersing it along a coarse-graining process.
The control of the effects of coarse-graining on the entropy production is indeed one major issue in the general context of statistical mechanics out of equilibrium~\cite{Seifert2019,Teza2020,Teza2020b}.

\section{Data availability}
The data used in this work are extracted from the BACI database \cite{CEPII:2010-23}, which is a refined version of the freely accessible COMDTRADE databse \cite{Comtrade2017} redacted by the United Nations. Reference tables of the Harmonized System 2007 nomenclature and ISO-3166 country codes can be found respectively at Ref. \cite{HS2007} and Ref. \cite{ISO3166}.

\bibliography{bibliography}

\begin{thebibliography}{10}
\urlstyle{rm}
\expandafter\ifx\csname url\endcsname\relax
  \def\url#1{\texttt{#1}}\fi
\expandafter\ifx\csname urlprefix\endcsname\relax\def\urlprefix{URL }\fi
\expandafter\ifx\csname doiprefix\endcsname\relax\def\doiprefix{DOI: }\fi
\providecommand{\bibinfo}[2]{#2}
\providecommand{\eprint}[2][]{\url{#2}}

\bibitem{Hidalgo2009}
\bibinfo{author}{Hidalgo, C.} \& \bibinfo{author}{Hausmann, R.}
\newblock \bibinfo{journal}{\bibinfo{title}{The building blocks of economic
  complexity}}.
\newblock {\emph{\JournalTitle{Proc. Natl. Acad. Soc. USA}}}
  \textbf{\bibinfo{volume}{106}}, \bibinfo{pages}{10570},
  \doiprefix\url{https://doi.org/10.1073/pnas.0900943106}
  (\bibinfo{year}{2009}).

\bibitem{Tacchella2012}
\bibinfo{author}{Tacchella, A.}, \bibinfo{author}{Cristelli, M.},
  \bibinfo{author}{Caldarelli, G.}, \bibinfo{author}{Gabrielli, A.} \&
  \bibinfo{author}{Pietronero, L.}
\newblock \bibinfo{journal}{\bibinfo{title}{A new metrics for countries'
  fitness and products' complexity}}.
\newblock {\emph{\JournalTitle{Sci. Rep.}}} \textbf{\bibinfo{volume}{2}},
  \bibinfo{pages}{723}, \doiprefix\url{https://doi.org/10.1038/srep00723}
  (\bibinfo{year}{2012}).

\bibitem{hausmann2014atlas}
\bibinfo{author}{Hausmann, R.} \emph{et~al.}
\newblock \emph{\bibinfo{title}{The atlas of economic complexity: Mapping paths
  to prosperity}} (\bibinfo{publisher}{Mit Press}, \bibinfo{year}{2014}).

\bibitem{Caraglio2016}
\bibinfo{author}{Caraglio, M.}, \bibinfo{author}{Baldovin, F.} \&
  \bibinfo{author}{Stella, A.~L.}
\newblock \bibinfo{journal}{\bibinfo{title}{Export dynamics as an optimal
  growth problem in the network of global economy}}.
\newblock {\emph{\JournalTitle{Sci. Rep.}}} \textbf{\bibinfo{volume}{6}},
  \doiprefix\url{https://doi.org/10.1038/srep31461} (\bibinfo{year}{2016}).

\bibitem{Teza2018}
\bibinfo{author}{Teza, G.}, \bibinfo{author}{Caraglio, M.} \&
  \bibinfo{author}{Stella, A.~L.}
\newblock \bibinfo{journal}{\bibinfo{title}{Growth dynamics and complexity of
  national economies in the global trade network}}.
\newblock {\emph{\JournalTitle{Scientific reports}}}
  \textbf{\bibinfo{volume}{8}}, \bibinfo{pages}{1--8},
  \doiprefix\url{https://doi.org/10.1038/s41598-018-33659-6}
  (\bibinfo{year}{2018}).

\bibitem{Teza2018b}
\bibinfo{author}{Teza, G.}, \bibinfo{author}{Caraglio, M.} \&
  \bibinfo{author}{Stella, A.~L.}
\newblock \bibinfo{journal}{\bibinfo{title}{Data driven approach to the
  dynamics of import and export of g7 countries}}.
\newblock {\emph{\JournalTitle{Entropy}}} \textbf{\bibinfo{volume}{20}},
  \bibinfo{pages}{735}, \doiprefix\url{https://doi.org/10.3390/e20100735}
  (\bibinfo{year}{2018}).

\bibitem{Helpman1991}
\bibinfo{author}{Grossman, G.} \& \bibinfo{author}{Helpman, E.}
\newblock \bibinfo{journal}{\bibinfo{title}{Quality ladders in the theory of
  growth}}.
\newblock {\emph{\JournalTitle{Rev. Econ. Stud.}}}
  \textbf{\bibinfo{volume}{58}}, \bibinfo{pages}{43},
  \doiprefix\url{https://doi.org/10.2307/2298044} (\bibinfo{year}{1991}).

\bibitem{Howitt:1998}
\bibinfo{author}{Aghion, P.} \& \bibinfo{author}{Howitt, P.}
\newblock \emph{\bibinfo{title}{Quality ladders in the theory of growth}}
  (\bibinfo{publisher}{MIT Press, Cambridge, MA}, \bibinfo{year}{1998}).

\bibitem{Kemp2014}
\bibinfo{author}{Kemp-Benedict, E.}
\newblock \bibinfo{journal}{\bibinfo{title}{An interpretation and critique of
  the method of reflections}}.
\newblock {\emph{\JournalTitle{Munich Personal RePEc Archive}}}
  \doiprefix\url{https://mpra.ub.uni-muenchen.de/60705/}
  (\bibinfo{year}{2014}).

\bibitem{Morrison2017}
\bibinfo{author}{Morrison, G.} \emph{et~al.}
\newblock \bibinfo{journal}{\bibinfo{title}{On economic complexity and the
  fitness of nations}}.
\newblock {\emph{\JournalTitle{Scientific Reports}}}
  \textbf{\bibinfo{volume}{7}}, \bibinfo{pages}{1--11},
  \doiprefix\url{https://doi.org/10.1038/s41598-017-14603-6}
  (\bibinfo{year}{2017}).

\bibitem{Servedio2018}
\bibinfo{author}{Servedio, V.}, \bibinfo{author}{Buttà, P.},
  \bibinfo{author}{Mazzilli, D.}, \bibinfo{author}{Tacchella, A.} \&
  \bibinfo{author}{Pietronero, L.}
\newblock \bibinfo{journal}{\bibinfo{title}{A new and stable estimation method
  of country economic fitness and product complexity}}.
\newblock {\emph{\JournalTitle{Entropy}}} \textbf{\bibinfo{volume}{20}},
  \bibinfo{pages}{783}, \doiprefix\url{10.3390/e20100783}
  (\bibinfo{year}{2018}).

\bibitem{Balassa1965}
\bibinfo{author}{Balassa, B.}
\newblock \bibinfo{journal}{\bibinfo{title}{Trade liberalisation and
  “revealed” comparative advantage 1}}.
\newblock {\emph{\JournalTitle{The manchester school}}}
  \textbf{\bibinfo{volume}{33}}, \bibinfo{pages}{99--123}
  (\bibinfo{year}{1965}).

\bibitem{Mealyeaau2019}
\bibinfo{author}{Mealy, P.}, \bibinfo{author}{Farmer, J.~D.} \&
  \bibinfo{author}{Teytelboym, A.}
\newblock \bibinfo{journal}{\bibinfo{title}{Interpreting economic complexity}}.
\newblock {\emph{\JournalTitle{Sci. Adv.}}} \textbf{\bibinfo{volume}{5}},
  \doiprefix\url{10.1126/sciadv.aau1705} (\bibinfo{year}{2019}).

\bibitem{Spellerberg2003}
\bibinfo{author}{Spellerberg, I.~F.} \& \bibinfo{author}{Fedor, P.~J.}
\newblock \bibinfo{journal}{\bibinfo{title}{A tribute to claude shannon
  (1916–2001) and a plea for more rigorous use of species richness, species
  diversity and the `shannon--wiener' index}}.
\newblock {\emph{\JournalTitle{Glob. Ecol. Biogeogr.}}}
  \textbf{\bibinfo{volume}{12}}, \bibinfo{pages}{177--179},
  \doiprefix\url{https://doi.org/10.1046/j.1466-822X.2003.00015.x}
  (\bibinfo{year}{2003}).

\bibitem{Jacquemin1979}
\bibinfo{author}{Jacquemin, A.~P.} \& \bibinfo{author}{Berry, C.~H.}
\newblock \bibinfo{journal}{\bibinfo{title}{Entropy measure of diversification
  and corporate growth}}.
\newblock {\emph{\JournalTitle{J. Ind. Econ.}}}
  \textbf{\bibinfo{volume}{XXVII}}, \bibinfo{pages}{359--369},
  \doiprefix\url{https://doi.org/10.2307/2097958} (\bibinfo{year}{1979}).

\bibitem{Saviotti2008}
\bibinfo{author}{Saviotti, P.~P.} \& \bibinfo{author}{Frenken, K.}
\newblock \bibinfo{journal}{\bibinfo{title}{Export variety and the economic
  performance of countries}}.
\newblock {\emph{\JournalTitle{J. Evol. Econ.}}} \textbf{\bibinfo{volume}{18}},
  \bibinfo{pages}{201--218}, \doiprefix\url{10.1007/s00191-007-0081-5}
  (\bibinfo{year}{2008}).

\bibitem{Shannon1948}
\bibinfo{author}{Shannon, C.~E.}
\newblock \bibinfo{journal}{\bibinfo{title}{A mathematical theory of
  communication}}.
\newblock {\emph{\JournalTitle{Bell Syst. Tech. J.}}}
  \textbf{\bibinfo{volume}{27}}, \bibinfo{pages}{379--423},
  \doiprefix\url{10.1002/j.1538-7305.1948.tb01338.x} (\bibinfo{year}{1948}).

\bibitem{Weaver1949}
\bibinfo{author}{Weaver, W.}
\newblock \bibinfo{journal}{\bibinfo{title}{The mathematics of communication}}.
\newblock {\emph{\JournalTitle{Scientific American}}}
  \textbf{\bibinfo{volume}{181}}, \bibinfo{pages}{11--15},
  \doiprefix\url{10.2307/24967225} (\bibinfo{year}{1949}).

\bibitem{Jost2006}
\bibinfo{author}{Jost, L.}
\newblock \bibinfo{journal}{\bibinfo{title}{Entropy and diversity}}.
\newblock {\emph{\JournalTitle{Oikos}}} \textbf{\bibinfo{volume}{113}},
  \bibinfo{pages}{363--375},
  \doiprefix\url{https://doi.org/10.1111/j.2006.0030-1299.14714.x}
  (\bibinfo{year}{2006}).

\bibitem{Teza2020b}
\bibinfo{author}{Teza, G.}
\newblock \emph{\bibinfo{title}{Out of equilibrium dynamics: from an entropy of
  the growth to the growth of entropy production}}.
\newblock Ph.D. thesis, \bibinfo{school}{University of Padova}
  (\bibinfo{year}{2020}).

\bibitem{Guillaume2004}
\bibinfo{author}{Guillaume, J.-L.} \& \bibinfo{author}{Latapy, M.}
\newblock \bibinfo{journal}{\bibinfo{title}{Bipartite structure of all complex
  networks}}.
\newblock {\emph{\JournalTitle{Information processing letters}}}
  \textbf{\bibinfo{volume}{90}}, \bibinfo{pages}{215--221},
  \doiprefix\url{https://doi.org/10.1016/j.ipl.2004.03.007}
  (\bibinfo{year}{2004}).

\bibitem{Newman2010}
\bibinfo{author}{Newman, M.}
\newblock \emph{\bibinfo{title}{{Networks: an introduction}}}
  (\bibinfo{publisher}{Oxford university press}, \bibinfo{year}{2010}).

\bibitem{Dormann2009}
\bibinfo{author}{Dormann, C.~F.}, \bibinfo{author}{Fr{\"u}nd, J.},
  \bibinfo{author}{Bl{\"u}thgen, N.} \& \bibinfo{author}{Gruber, B.}
\newblock \bibinfo{journal}{\bibinfo{title}{Indices, graphs and null models:
  analyzing bipartite ecological networks}}.
\newblock {\emph{\JournalTitle{The Open Ecology Journal}}}
  \textbf{\bibinfo{volume}{2}},
  \doiprefix\url{http://doi.org/10.2174/1874213000902010007}
  (\bibinfo{year}{2009}).

\bibitem{Chacoff2012}
\bibinfo{author}{Chacoff, N.~P.} \emph{et~al.}
\newblock \bibinfo{journal}{\bibinfo{title}{Evaluating sampling completeness in
  a desert plant--pollinator network}}.
\newblock {\emph{\JournalTitle{Journal of Animal Ecology}}}
  \textbf{\bibinfo{volume}{81}}, \bibinfo{pages}{190--200},
  \doiprefix\url{https://doi.org/10.1111/j.1365-2656.2011.01883.x}
  (\bibinfo{year}{2012}).

\bibitem{Allesina2008}
\bibinfo{author}{Allesina, S.} \& \bibinfo{author}{Pascual, M.}
\newblock \bibinfo{journal}{\bibinfo{title}{Network structure, predator--prey
  modules, and stability in large food webs}}.
\newblock {\emph{\JournalTitle{Theoretical Ecology}}}
  \textbf{\bibinfo{volume}{1}}, \bibinfo{pages}{55--64},
  \doiprefix\url{https://doi.org/10.1007/s12080-007-0007-8}
  (\bibinfo{year}{2008}).

\bibitem{Ramasco2004}
\bibinfo{author}{Ramasco, J.~J.}, \bibinfo{author}{Dorogovtsev, S.~N.} \&
  \bibinfo{author}{Pastor-Satorras, R.}
\newblock \bibinfo{journal}{\bibinfo{title}{Self-organization of collaboration
  networks}}.
\newblock {\emph{\JournalTitle{Physical review E}}}
  \textbf{\bibinfo{volume}{70}}, \bibinfo{pages}{036106},
  \doiprefix\url{https://doi.org/10.1103/PhysRevE.70.036106}
  (\bibinfo{year}{2004}).

\bibitem{Corel2018}
\bibinfo{author}{Corel, E.} \emph{et~al.}
\newblock \bibinfo{journal}{\bibinfo{title}{Bipartite network analysis of gene
  sharings in the microbial world}}.
\newblock {\emph{\JournalTitle{Molecular biology and evolution}}}
  \textbf{\bibinfo{volume}{35}}, \bibinfo{pages}{899--913},
  \doiprefix\url{https://doi.org/10.1093/molbev/msy001} (\bibinfo{year}{2018}).

\bibitem{CEPII:2010-23}
\bibinfo{author}{Gaulier, G.} \& \bibinfo{author}{Zignago, S.}
\newblock \bibinfo{title}{Baci: International trade database at the
  product-level. the 1994-2007 version}.
\newblock \bibinfo{type}{Working Papers} \bibinfo{number}{2010-23},
  \bibinfo{institution}{CEPII} (\bibinfo{year}{2010}).
\newblock \doiprefix\url{https://dx.doi.org/10.2139/ssrn.1994500}.

\bibitem{Comtrade2017}
\bibinfo{author}{UN}.
\newblock \emph{\bibinfo{title}{Commodity Trade Statistics Database}}
  (\bibinfo{year}{2017 (accessed June 24, 2020)}).
\newblock \bibinfo{note}{\url{https://comtrade.un.org}}.

\bibitem{HS2007}
\bibinfo{author}{WCO}.
\newblock \emph{\bibinfo{title}{Harmonic System Nomenclature}}
  (\bibinfo{year}{2007 (accessed June 24, 2020)}).
\newblock \bibinfo{note}{\url{http://www.wcoomd.org}}.

\bibitem{ISO3166}
\bibinfo{author}{ISO}.
\newblock \emph{\bibinfo{title}{ISO 3166 Country Codes}} (\bibinfo{year}{2021
  (accessed June 24, 2020)}).
\newblock
  \bibinfo{note}{\url{https://https://www.iso.org/iso-3166-country-codes.html}}.

\bibitem{Kellogg1976}
\bibinfo{author}{Kellogg, R.~B.}, \bibinfo{author}{Li, T.-Y.} \&
  \bibinfo{author}{Yorke, J.}
\newblock \bibinfo{journal}{\bibinfo{title}{A constructive proof of the brouwer
  fixed-point theorem and computational results}}.
\newblock {\emph{\JournalTitle{SIAM Journal on Numerical Analysis}}}
  \textbf{\bibinfo{volume}{13}}, \bibinfo{pages}{473--483},
  \doiprefix\url{https://doi.org/10.1137/0713041} (\bibinfo{year}{1976}).

\bibitem{Evans2015}
\bibinfo{author}{Evans, L.~C.} \& \bibinfo{author}{Gariepy, R.~F.}
\newblock \emph{\bibinfo{title}{Measure theory and fine properties of
  functions}} (\bibinfo{publisher}{CRC press}, \bibinfo{year}{2015}).

\bibitem{Hogg:1995}
\bibinfo{author}{Hogg, R.} \& \bibinfo{author}{Craig, A.}
\newblock \emph{\bibinfo{title}{Introduction to Mathematical Statistics}}
  (\bibinfo{publisher}{New York: Macmillan}, \bibinfo{year}{1995}).

\bibitem{Seifert2019}
\bibinfo{author}{Seifert, U.}
\newblock \bibinfo{journal}{\bibinfo{title}{From stochastic thermodynamics to
  thermodynamic inference}}.
\newblock {\emph{\JournalTitle{Annual Review of Condensed Matter Physics}}}
  \textbf{\bibinfo{volume}{10}}, \bibinfo{pages}{171--192},
  \doiprefix\url{10.1146/annurev-conmatphys-031218-013554}
  (\bibinfo{year}{2019}).

\bibitem{Teza2020}
\bibinfo{author}{Teza, G.} \& \bibinfo{author}{Stella, A.~L.}
\newblock \bibinfo{journal}{\bibinfo{title}{Exact coarse graining preserves
  entropy production out of equilibrium}}.
\newblock {\emph{\JournalTitle{Phys. Rev. Lett.}}}
  \textbf{\bibinfo{volume}{125}}, \bibinfo{pages}{110601},
  \doiprefix\url{10.1103/PhysRevLett.125.110601} (\bibinfo{year}{2020}).

\end{thebibliography}

\section*{Acknowledgments}

G.T. is supported by the Center for Statistical Mechanics at the Weizmann Institute of Science and by the grant 662962 of of the Simons foundation and the grant 873028 of the EU Horizon 2020 program. All the authors thank the Physics and Astronomy Department of Padua University for financial support.

\section*{Author contributions statement}

G.T. collected and organized the data and carried out the numerical analyses. G.T., M.C. and A.L.S. devised the research, analyzed the data and wrote the manuscript.

\section*{Additional information}

\textbf{Competing interests:} The authors declare no competing interests.

\end{document}